\documentclass[sigconf]{acmart-me}

\usepackage{booktabs} 
\usepackage{url}
\usepackage{color}
\usepackage{multirow}
\usepackage{enumitem}
\usepackage{array}
\usepackage{siunitx}
\usepackage[labelfont={bf}]{caption}
\usepackage{balance}
\usepackage{array} 

\acmDOI{}
\acmISBN{}
\acmConference[MediaEval'20]{Multimedia Evaluation Workshop}{December 14-15 2020}{Online}
\copyrightyear{}
\acmYear{}
\acmPrice{}

\begin{document}
\title{Automatic Polyp Segmentation using Fully Convolutional Neural Network}

\author{Nikhil Kumar Tomar\textsuperscript{1}}
\affiliation{
\textsuperscript{1}Indira Gandhi National Open University, India
}
\email{nikhilroxtomar@gmail.com}

\renewcommand{\shorttitle}{MediaEval'20: Multimedia Evaluation Workshop}
\renewcommand{\shortauthors}{N. K. Tomar et. al.}

\begin{abstract}
Colorectal cancer is one of fatal cancer worldwide. Colonoscopy is the standard treatment for examination, localization, and removal of colorectal polyps.  However, it has been shown that the miss-rate of colorectal polyps during colonoscopy is between 6 to 27\%~\cite{ahn2012miss}. The use of an automated, accurate, and real-time polyp segmentation during colonoscopy examinations can help the clinicians to eliminate missing lesions and prevent further progression of colorectal cancer. The ``Medico automatic polyp segmentation challenge'' provides an opportunity to study polyp segmentation and build a fast segmentation model. The challenge organizers provide a Kvasir-SEG dataset to train the model. Then it is tested on a separate unseen dataset to validate the efficiency and speed of the segmentation model. The experiments demonstrate that the model trained on the Kvasir-SEG dataset~\cite{jha2020kvasir} and tested on an unseen dataset achieves a dice coefficient of 0.7801, mIoU of 0.6847, recall of 0.8077, and precision of 0.8126, demonstrating the generalization ability of our model. The model has achieved 80.60 FPS on the unseen dataset with an image resolution of $512 \times 512$.
\vspace{-4.5mm}
\end{abstract}

\maketitle
\section{Introduction}
\label{sec:intro}
Colorectal cancer is one of the dangerous types of cancer, adding to significant deaths worldwide.  Polyps are an early indicator of this type of cancer, and clinicians often detect it through colonoscopy. These polyps come in various shapes and sizes and are sometimes missed by clinicians as some polyps are hard to differentiate from the surrounding tissue. Sometimes these polyps are covered with stool, mucosa, and other surrounding structures and pose challenges for clinicians. This is why it is essential to build a Computer-Aided Diagnosis (CADx) system for detecting polyps.

The automatic polyp segmentation can play a crucial role in identifying and localizing the affected regions from the images or video frames. Semantic segmentation helps you analyze each pixel and classify them into a well-defined polyp or non-polyp class instance. With the increase in the amount of publicly available datasets, dominant methodology such as convolutional neural networks and improved hardware enables researchers to solve the challenging task of automated diagnosis of colorectal cancer in real-time.

The ``Medico Automatic Polyp Segmentation Challenge''~\cite{jha2020medico} consists of two tasks. The first task is ``Polyp segmentation task'' and the second is ``Algorithm efficiency task''. A single has been submitted for both the task. The model is efficient in terms of both the evaluation metrics score and the FPS.

\begin{figure*}
 \center
  \includegraphics[height=10cm]{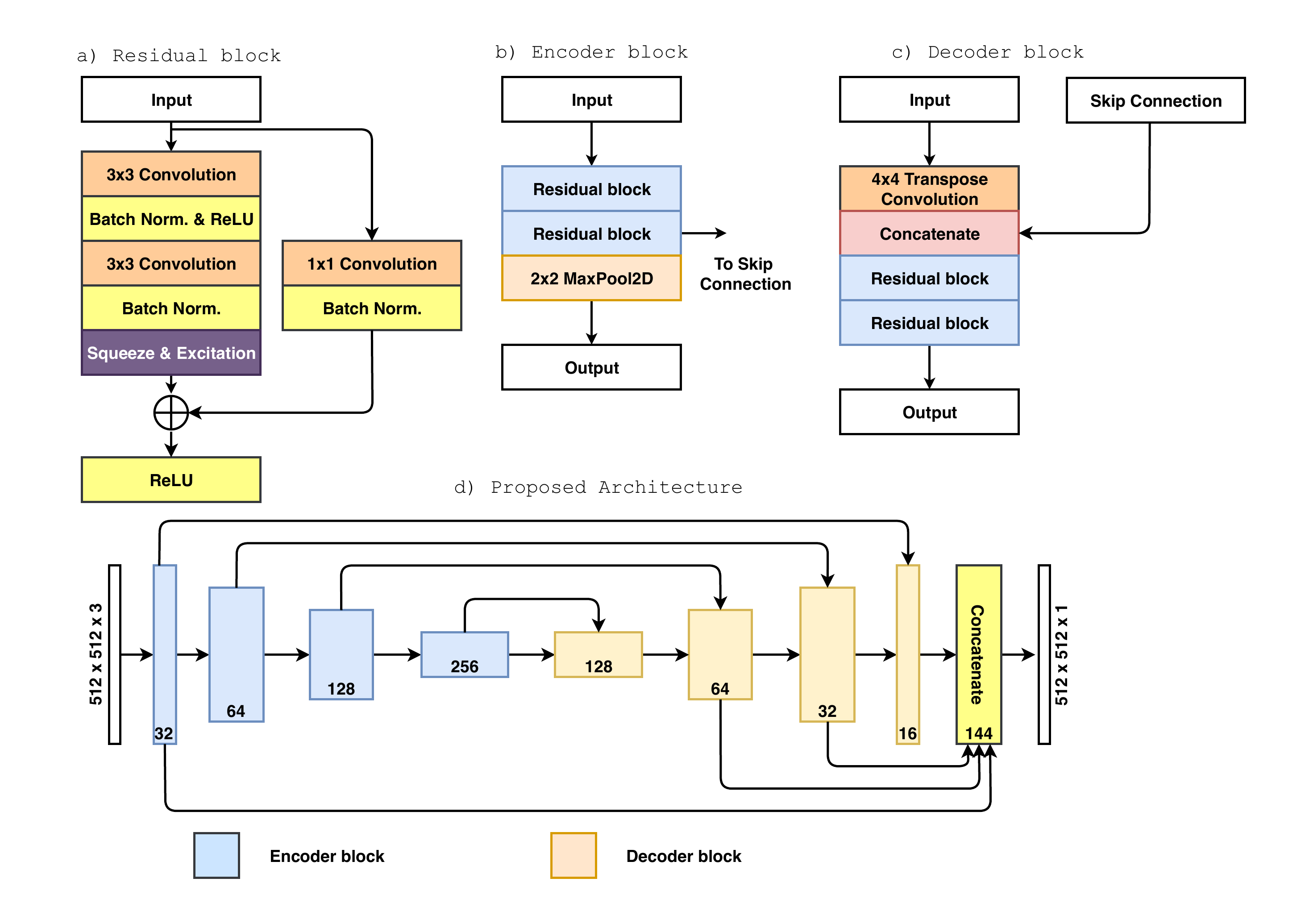}
  \vspace{-0.5cm}
  \caption{The proposed architecture and its components}
  \label{fig:architecture_fig}
\end{figure*}

\vspace{-3mm}

\section{Approach}
\label{sec:approach}
The proposed architecture is a full convolution network following an encoder-decoder approach. It combines the strength of residual learning~\cite{he2016deep} and the attention mechanism of the squeeze and excitation network~\cite{hu2018squeeze}. The encoding network consists of $4$ encoder block with $32$, $64$, $128$, and $256$ number of filters. The decoding network also consists of a $4$ decoder block with $128$, $64$, $32$, and $16$ number of filters. Both the encoder and decoder block consists of a residual block as their core component.

The residual block consists of two $3 \times 3$ convolutions, where the first convolution is followed by a Batch Normalization (BN) and a Rectified Linear Unit (ReLU) activation function. Next comes the second convolution layer with a BN and a squeeze and excitation layer, which is then added with the residual block's input. After that, a ReLU activation function is used, which acts as the output of the residual block. The residual block helps us in building deep neural networks by solving the vanishing gradient and exploding gradient problem.

The convolution operation takes an image to apply a number of filters generating a new set of output features maps. The squeeze and excitation layer provides channel-wise attention to these feature maps, strengthening the important feature and suppressing irrelevant features. The squeeze and excitation is two steps approach. At first, the incoming feature maps are first squeezed by using a global average pooling operation. This helps to get a global feature vector of size $n$, where $n$ represents the number of feature channels from the incoming feature maps. In the second step(excitation), this global feature vector goes through a two-layer feed-forward neural network. Here the number of features is first reduced and then expanded to the original size $n$. Finally, a sigmoid activation function is used, which scales the feature vector value between $0$ and $1$. This scaled feature vector is used to multiply the incoming feature maps.

The proposed network takes the polyp image of $512 \times 512$ size as the input given to the first encoder block. Each encoder block starts with two residual blocks, where each residual block consists of two $3 \times 3$ convolution and a shortcut connecting the output of both the convolution layer, also called the identity mapping. The output of the second residual block acts as the skip connection of the corresponding decoder block. These skip connections provide early information to the decoder block, which improves feature reconstruction and better model performance. It is followed by a $2 \times 2$ max-pooling operation, which downscales the incoming feature maps. This process is repeated $4$ times, thereby reducing the feature maps size by $8$ times ($64 \times 64$) the original input size. The output of the last encoder block acts as the input of the first decoder block. In each decoder block, first, the incoming feature map is upscaled by using a $4 \times 4$ transpose convolution. After that, the upscaled feature map is concatenated with the appropriate same resolution feature map from the encoder via skip-connection. The skip connection provides information that is sometimes lost due to the depth of the network. It helps the decoder in the better reconstruction of the semantic feature maps. Next, follows the two residual blocks assisting the network in to learn necessary features via back-propagation. This process is repeated four times, thereby getting the output feature map of size $512 \times 512$. 
Now, the last three decoder blocks' output is taken and upscaled to the resolution of $512 \times 512$ using a $4 \times 4$ transpose convolution. Next, we concatenate these three upscaled feature maps and the skip-connection feature maps from the first encoder block. These concatenated feature maps are then passed through a $1 \times 1$ convolution and sigmoid activation function. The output of the sigmoid activation results in a binary mask.

\section{Results and Analysis}
\label{sec:results}

Table~\ref{table:quantative_results_task1} shows the overall results on the validation dataset of Kvasir-SEG and unseen test dataset provided by the challenge organizers. For the evaluation of the results, the Mean Intersection-Over-Union (mIoU), Sørensen–Dice coefficient (DSC), recall, precision (Prec.), accuracy (Acc.), and F2 metrics were used for both task 1 and 2. Additionally, FPS was also calculated for task 2. Task 1 and task 2's evaluation score is the same as a single model was used for both the tasks. The proposed model trained on the Kvasir-SEG dataset~\cite{jha2020kvasir} and tested on an unseen dataset achieves a DSC of $0.7801$, mIoU of $0.6847$, recall of $0.8077$, precision of $0.8126$, accuracy of $0.9404$, and F2 of $0.7854$, demonstrating the generalization ability. The model has achieved 80.60 FPS on the unseen test dataset with an image resolution of $512 \times 512$. This shows that the model is capable of giving competitive results with higher input image resolution in real-time.

\begin{table}[t]
 \caption{Quantitative results on Kvasir-SEG and unseen (Challenge) dataset for task 1 and 2.}
    \label{table:quantative_results_task1}
    \def\arraystretch{1.5}
     \setlength\tabcolsep{2pt}
    \centering
  \begin{tabular}{@{}l|l|l|l|l|l|l|l@{}}
\hline
\textbf{Dataset}& \textbf{mIoU} & \textbf{DSC} & \textbf{Recall} & \textbf{Prec.} & \textbf{Acc.} & \textbf{F2} & \textbf{FPS} \\ \hline
Kvasir-SEG & 0.7565 & 0.8411 & 0.8643 & 0.8680 & 0.9532 & 0.8461 & - \\ \hline
Unseen     & 0.6847 & 0.7801 & 0.8077 & 0.8126 & 0.9404 & 0.7854 & 80.60\\ \hline
\end{tabular}
\vspace{-8mm}
\end{table} 

\section{Conclusion}
\label{sec:conclusion}
The Medico Automatic Polyp Segmentation challenge~\cite{jha2020medico} provides a platform to explore the potential and challenges of automated polyp segmentation on the Kvasir-SEG dataset containing 1000 images and their respective annotative masks. We have trained the proposed model and provide competitive results for both task 1 and task 2. We believe this approach will be an effective method for the rapid and automated segmentation of polyps. In the future, we can further investigate how to improve the system by further reducing the model complexity while improving performance. 

\bibliographystyle{ACM-Reference-Format}
\def\bibfont{\small} 
\balance
\bibliography{sigproc} 

\end{document}